\documentclass[12pt,a4paper]{article}

  \usepackage{amsmath}
  \usepackage{graphicx}

  \usepackage{hyperref} 
  \pdfoutput=1

  \title{Rayleigh-Taylor instability between two stable stratifications}
  \author{Megan S. Davies Wykes and Stuart B. Dalziel \\ \normalsize DAMTP, University of Cambridge, UK \\ \small M.S.Davies-Wykes@damtp.cam.ac.uk}
  \date{}
  
\begin{document}

  \maketitle

\vspace{-0.5cm}

These fluid dynamics video sequences show two Rayleigh-Taylor instability experiments. The first video sequence shows an experiment where two layers of uniform density are arranged such that the density of the upper layer is greater than the density of the lower. The unstable interface between the two layers is initially supported by a stainless steel barrier.  With the removal of the barrier, Rayleigh-Taylor instability results in the creation of a mixing region, which grows with time. This growth accelerates until the mixing region fills the entire tank.

\begin{figure}[ht]
                \centering
                \includegraphics[width=0.15\textwidth]{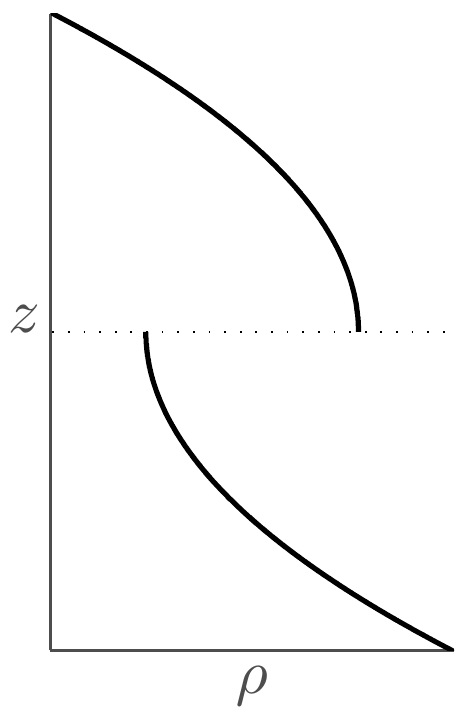}
                \caption{Sketch of the initial density profile present in the tank.}
                \label{fig:basic}
        \end{figure}

The second sequence shows Rayleigh-Taylor instability when it is confined between two stable stratifications. A sketch of the density profile at the start of this experiment is shown in figure \ref{fig:basic}. Although initially the growth of the mixing region accelerates, the stable stratification slows the growth of the instability and brings it to a halt before it can fill the tank.

The stratifications are constructed of fresh and salt water and both experiments have the same Atwood number, $A = (\rho_{\text{upper}} - \rho_{\text{lower}})/(\rho_{\text{upper}} + \rho_{\text{lower}}) = 7 \times 10^{-4}$, where $\rho_{\text{upper}}$ and $\rho_{\text{lower}}$ are the densities just above and just below the barrier at the start of the experiment. Visualisation of the upper surface of the mixing layer was achieved by the use of fluorescent dye in the lower layer. The tank used measures $0.5$m $\times \, 0.4$m $\times \, 0.2$m.

  \end{document}